# Ultra-Wide Bandgap AlGaN Heterostructure Field Effect Transistors with Current Gain Cutoff Frequency Above 85 GHz

Yinxuan Zhu[1, a], Andrew A. Allerman[3], Ashley Wissel-Garcia[4], Seungheon Shin[1], Jonathan Pratt[1], Can Cao[1], Kyle J. Liddy[1,5], James S. Speck[4], Brianna A. Klein[3], Andrew Armstrong[3], Siddharth Rajan[1,2,a]

[1] *Department of Electrical and Computer Engineering, The Ohio State University, Columbus, Ohio 43210, USA*

[2] *Department of Materials Science and Engineering, The Ohio State University, Columbus, Ohio 43210, USA*

[3] *Sandia National Laboratories, Albuquerque, New Mexico 87123, USA.*

[4] *Materials Department, University of California, Santa Barbara, California 93106, USA.*

[5] *Air Force Research Laboratory, Sensors Directorate, Wright-Patterson Air Force Base, Ohio 45433, USA*

**Abstract:**

We report the design and demonstration of ultra-wide-bandgap (UWBG) AlGaN polarization-graded field-effect transistors (PolFETs) that achieve a current-gain cutoff frequency above 85 GHz and a current density exceeding 1.3 A/mm. Ultra-thin channel and buffer layers were grown epitaxially on AlN substrates, and a reverse-graded AlGaN contact layer was incorporated to reduce the contact resistance to below 1 Ω·mm. With aggressively scaled device dimensions, the AlGaN PolFETs exhibit state-of-the-art high-frequency performance for UWBG transistors. Small-signal modeling reveals both parasitic and transit delays, confirming the benefits of reduced access resistance and enhanced intrinsic transconductance. These results establish a new performance benchmark for UWBG AlGaN devices and demonstrate their strong potential for next-generation millimeter-wave electronics.

[a] Authors to whom correspondence should be addressed

Electronic mail: *zhu.2931@osu.edu, rajan.21@osu.edu*

Ultra-wide bandgap (UWBG) AlGaN is a promising candidate to surpass GaN lateral transistors for next-generation mm-wave power amplifiers if its high critical field can be harnessed to enable higher Johnson Figure of Merit [1,2]. In recent years, remarkable improvements were achieved on UWBG AlGaN-based devices [3-6], with breakdown field of over 8.5 MV/cm in AlGaN heterojunction diode with high-k dielectric [7], and average breakdown fields over 5 MV/cm breakdown field and 400 mA/mm current density in UWBG AlGaN heterostructure field effect transistors (HFETs) [8]. Furthermore, promising RF performances with $f_T/f_{max}$ = 40 GHz/58 GHz and current density over 400 mA/mm have been reported on UWBG AlGaN high electron mobility field effect transistors (HEMTs) [9]. The challenges involved in the design of UWBG AlGaN transistors are distinct from those of conventional GaN-channel transistors in terms of contact/channel design and buffer/epitaxial growth. The cutoff frequency and transconductance for state-of-the-art AlGaN transistors remain significantly lower than those of GaN-based devices due to a combination of factors, including the high contact resistance that limits current density and introduces large parasitic delays, degrading RF performance [10].

Various strategies have been explored to mitigate this issue, including reverse graded AlGaN contacts [11] and regrown contacts [6]. The integration of reverse-graded n++ AlGaN contact layers with abrupt channel heterostructure field effect transistors is challenging. When an *ex-situ* regrown approach is used to integrate contacts, interfacial layers between regrowth and channel [12] can degrade contact resistance. When *in-situ grown* reverse graded contact layers on HEMTs are used, heterojunction barriers between the barrier layer and channel layer limit the injection of carriers into the contacts [13]. Graded channel transistors present an elegant approach to integrate reverse-graded contacts with a channel since they eliminate any band offsets that would create barriers between the contact region and the channel region. Polarization-graded field effect transistors (PolFETs) integrated with reverse graded contact layers with Al-composition < 50% were demonstrated previously [14]. Graded polarization-induced channels have several advantages for RF device operation since the electron charge is spatially distributed that can improve electron velocity and linearity [15-17].

In addition to current injection challenges, the low thermal conductivity of AlGaN [18] also necessitates careful design for thermal management. A thin AlGaN buffer/channel grown on AlN has advantages both from the point of view of pseudomorphic growth as well as from the thermal resistance. AlN bulk substrates serve as excellent heat-spreading layers for UWBG AlGaN transistors. Thin AlGaN epitaxial channel/buffer layers must be designed to mitigate the impact of the negative polarization charge between AlN/AlGaN channel due to the discontinuity in polarization at the AlN/AlGaN interface. Previous demonstrations of AlGaN-channel transistors [3-6, 8-10, 19-20] use thick AlGaN buffer layers as the transition layer between AlN and AlGaN channel, but this results in both additional thermal resistance due to poor thermal conductivity of AlGaN alloys, as well as partial or complete relaxation in epitaxial structures. Therefore, in this work, we use ultra-thin UWBG AlGaN graded channel layers with a Si-delta doped back barrier to compensate the negative polarization.

The epitaxial design used here, shown in Figure 1 (a), was grown on an AlN substrate by metalorganic chemical vapor deposition (MOCVD) and consists of an AlN buffer terminated with Si delta-doping to compensate for negative polarization charge, 2 nm AlN, 0.5 nm UID $Al_{0.5}Ga_{0.5}N$ spacer layer, a 10 nm UID AlGaN *channel layer* graded from $Al_{0.5}Ga_{0.5}N$ to $Al_{0.8}Ga_{0.2}N$, a 20 nm n-$Al_{0.8}Ga_{0.2}N$ cap layer to compensate for surface depletion, a 25 nm heavily doped AlGaN *contact layer* graded from $Al_{0.82}Ga_{0.18}N$ to $Al_{0.14}Ga_{0.86}N$, and a 7.5 nm n++ $Al_{0.14}Ga_{0.86}N$ cap layer. Si delta-doping as the transition between buffer and spacer layers was carefully designed to compensate for negative polarization charge at $Al_{0.5}Ga_{0.5}N$/AlN interface while avoiding a parasitic electron channel. The layer thicknesses and Al composition were confirmed using high resolution X-ray diffraction (HR-XRD) (Bruker XRD) measurements, and fully strained growth was confirmed using XRD-reciprocal space mapping (RSM) measurements shown in Fig. 2(a) and 2(b).

The process flow started with ohmic metal patterning and evaporation of Ti/Al/Ni/Au (20 nm/120 nm/30 nm/100 nm), followed by $Cl_2$-based inductively-coupled-plasma reactive-ion etch (ICP-RIE) for mesa isolation. The recess etch region was patterned by electron beam lithography and done using a low-power $Cl_2$-based ICP-RIE etch process. After defining the access region by recess etching, gates were patterned by electron beam lithography and deposited with Ni/Au (20 nm/100 nm) using electron beam evaporation.

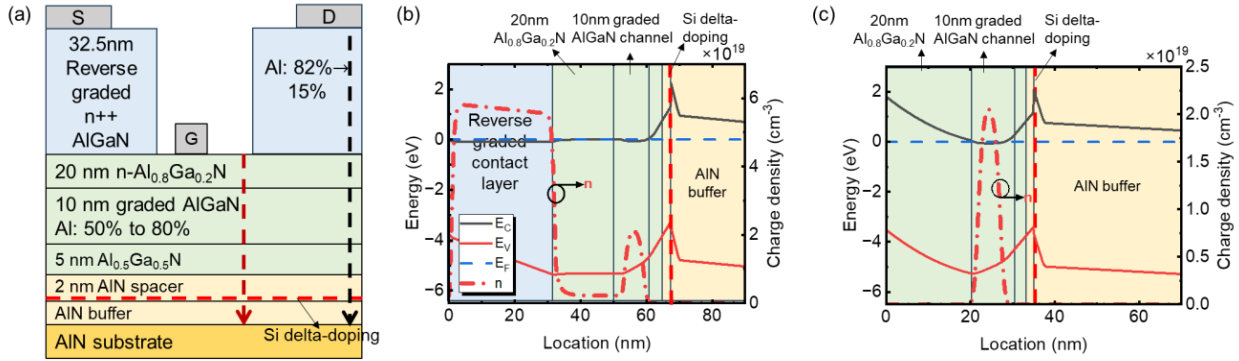

figure 1 (a) Schematic of UWBG AlGaN PolFETs; 1-D Schrodinger simulation of band diagrams, electron charge profile indicated by (b) black arrow; and (c) red arrow.

A sheet charge density of $1 \times 10^{13}$ cm$^{-2}$, electron mobility of 201 cm$^2$/V.s, and sheet resistivity of 2.97 kΩ/□ were extracted from Hall measurements on van der Pauw structures. DC characteristics were measured using an Agilent B1500 parameter analyzer. As shown in figure 3(a), contact resistance of 1 Ω.mm/$3.42 \times 10^{-6}$ Ω.cm$^2$ was extracted from transfer length measurements (TLM), meaning that our reverse graded contact layer significantly reduced the contact resistance on this UWBG AlGaN channel. Capacitance-voltage measurements were used to extract charge density dependence on depth in the access region (Fig. 3(c)). The plot of charge density versus depth matches well with the simulated one. This indicates Si delta-doping can compensate for the negative polarization charge at the AlGaN/AlN interface and simultaneously avoid a parallel electron channel.

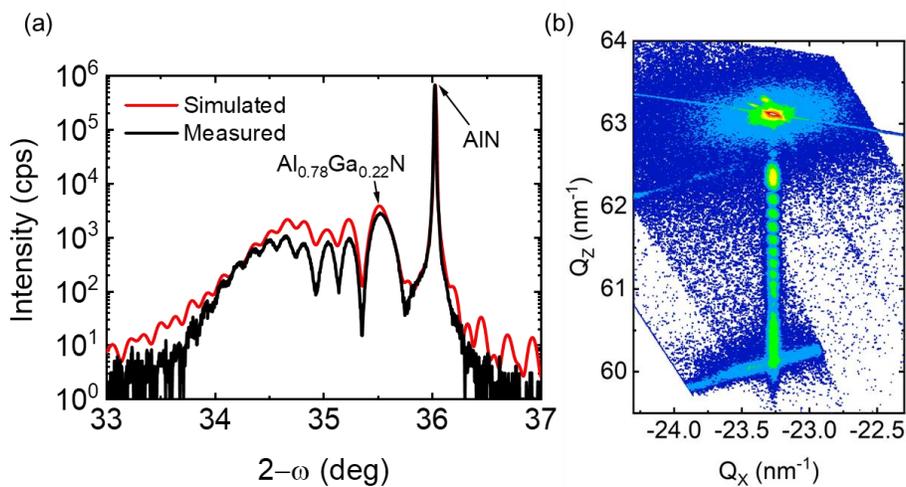

Figure 2 (a) XRD 2θ-ω scan of PolFET epitaxial layers; (b) XRD reciprocal space mapping data of the epitaxial layer stack.

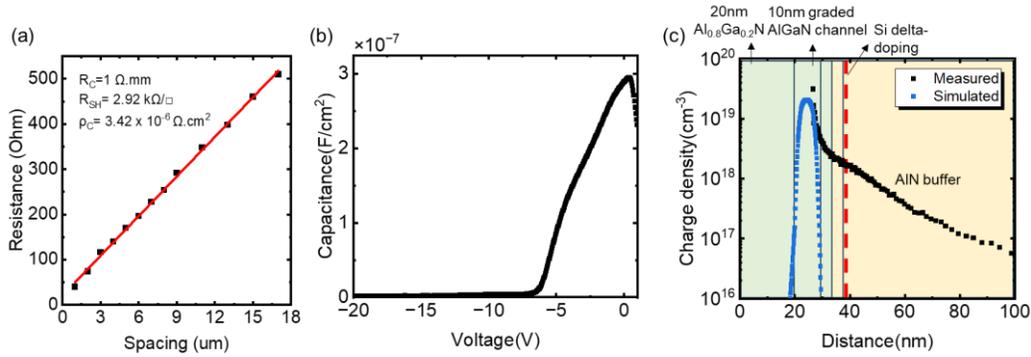

Figure 3 (a) TLM results; (b) Capacitance-voltage (CV) measurements on gate diode of AlGaN PolFET; (c) Extracted charge profile from CV profile and simulated charge profile in Figure 1(c).

Scaled transistors are shown in Fig. 4(a) with source-drain spacing of 200 nm, gate-drain spacing of 60 nm and gate length of 80 nm. A maximum current density of over 1.3 A/mm was measured at $V_{GS}$ = 3 V and $V_{DS}$ = 10 V, as shown in Fig. 4(b). This is the highest current density value obtained in planar UWBG AlGaN transistors. High current density was attributed to reduced contact resistance and aggressively scaled source-drain spacing. A transconductance of approximately 130 mS/mm and threshold voltage of -10 V were extracted from transfer characteristics (Fig. 4(c)). The on-off ratio ($\sim 10^4$) as shown in figure 4(d) was limited by the high gate leakage, and the 3-terminal breakdown voltage on similar devices was found to be 20 V, corresponding to an average field of 3.3 MV/cm. Future iterations could include dielectric layers to reduce the leakage and improve breakdown voltage, as demonstrated previously [4].

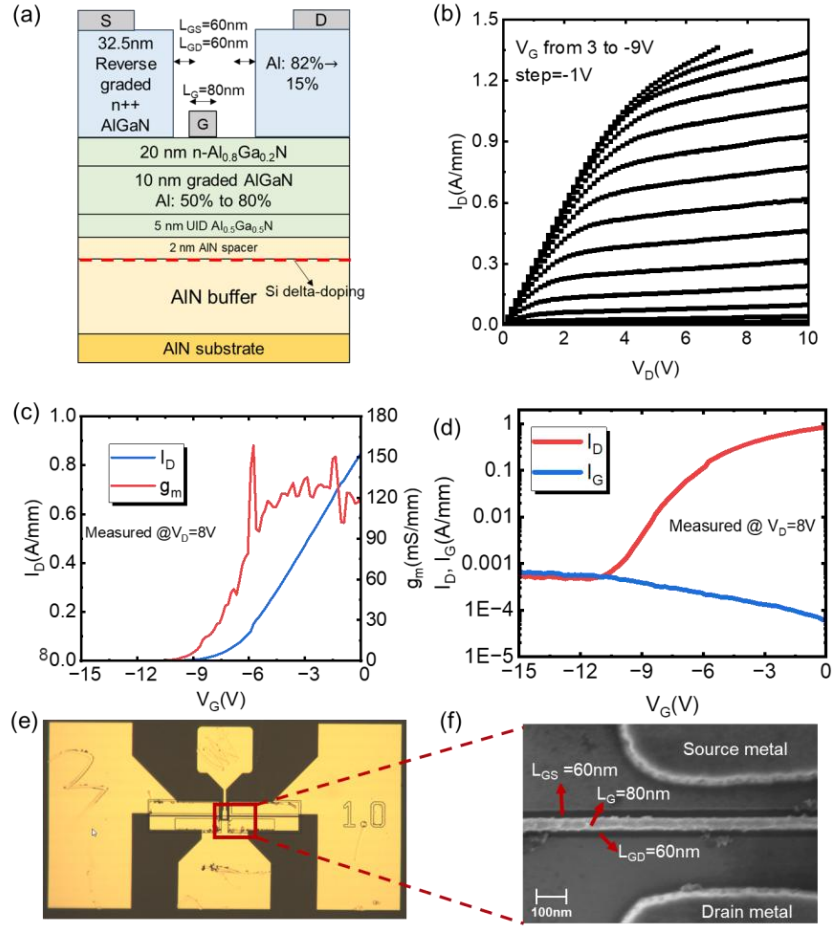

Figure 4 (a) Device Schematic of AlGaN PolFETs; (b) Output characteristics; (c) Transfer characteristics; (d) Drain/gate current density versus gate voltage in log scale; (e) Optical image of the device; (f) Top view SEM image of the region highlighted by the red square in (e).

On-wafer, small signal measurements were taken using an Agilent network analyzer 8722ES. The on-wafer short-circuit current gain (|h21|), maximum unilateral gain (MUG) and maximum stable gain (MSG) with gate biased at -3 V and drain biased at 10 V are shown in Fig. 5. We obtained a current gain cutoff frequency ($f_T$) of 85 GHz and maximum oscillation frequency ($f_{max}$) of 20 GHz. This is the record high cutoff frequency ($f_T$) for any UWBG AlGaN transistors. The relatively low $f_{max}/f_T$ ratio is due to high gate resistance from the i-shaped gate geometry as no T-shaped gate was used.

Small-signal equivalent circuit modeling was used to further analyze the RF performance of the transistor. The equivalent circuit model were extracted directly from small-signal measurements using a previously developed method [20], and the parasitic delay and transit delay of the transistor were calculated [21] to be 0.93 ps and 1.03 ps, respectively. These results indicate that by reducing the contact resistance and aggressively scaling the transistor dimensions, the source/drain access resistances can be substantially lowered, thereby enhancing the intrinsic transconductance and effectively minimizing both parasitic and transit delays.

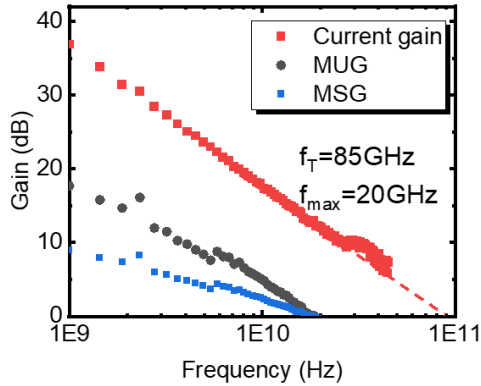

Figure 5 On-wafer small signal measurements.

Table 1. Elements extracted from S-parameters in small signal equivalent circuits.

| Intrinsic parameters | |
|---|---|
| $C_{gd}$ | 5.2 x 10$^{-14}$ F/mm |
| $C_{gs}$ | 1.45 x 10$^{-13}$ F/mm |
| $g_{m,int}$ | 141 mS/mm |
| $R_{ds}$ | 7.6 Ω.mm |
| **Extrinsic parameters** | |
| $R_s$ | 1.2 Ω.mm |
| $R_d$ | 1.2 Ω.mm |
| $R_g$ | 8 Ω/mm |
| **Delays** | |
| $\tau_{tran}$ | 1.03 ps |
| $\tau_p$ | 0.93 ps |

In conclusion, we designed and demonstrated a UWBG AlGaN PolFET with high current density of over 1.3 A/mm and current gain cutoff frequency ($f_T$) of 85 GHz (figure 6(a)(b)). The achieved current density is close to the theoretical limit calculated using the equation proposed in [22], which is attributed to the low contact resistance and scaled source–drain spacing. In addition, the combination of low contact resistance and reduced gate length enable state-of-art values for cutoff-frequency among AlGaN devices. The ideas demonstrated in this PolFET structure establish a promising design platform for future UWBG AlGaN-based power amplifier with low contact resistance, good electron transport and excellent thermal management.

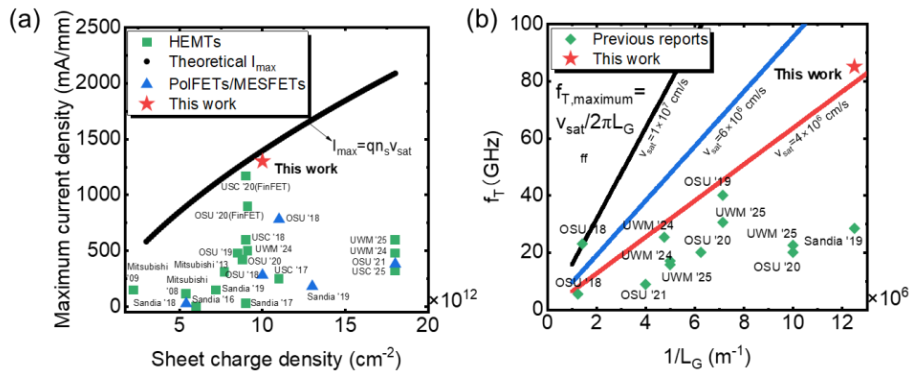

Figure 6 Benchmark of (a) maximum current density ($I_{max}$) as a function of sheet charge density ($n_s$). Also shown is the theoretical maximum current density in GaN following Bajaj et al [22]. (b)

current gain cutoff frequency (f$_T$) as a function of inverse gate length (1/L$_G$), and theoretical intrinsic current gain cutoff frequency limits based on gate transit time ($L_g/v_{sat}$).